\documentstyle[psfig,11pt]{article}
\def\gsim{\;
\raise0.3ex\hbox{$>$\kern-0.75em\raise-1.1ex\hbox{$\sim$}}\;}
\def\lsim{\;
\raise0.3ex\hbox{$<$\kern-0.75em\raise-1.1ex\hbox{$\sim$}}\;g}

\newcommand{\beq}{\begin{equation}}
\newcommand{\eeq}{\end{equation}}

\def\lsim{\raise0.3ex\hbox{$\;<$\kern-0.75em\raise-1.1ex\hbox{$\sim\;$}}}
\def\gsim{\raise0.3ex\hbox{$\;>$\kern-0.75em\raise-1.1ex\hbox{$\sim\;$}}}

\def\nt{\hbox{$\nu_\tau$ }}

\def\apj#1#2#3{          {\it Astrophys. J. }{\bf #1} (19#2) #3}

\def\nat#1#2#3{          {\it Nature }{\bf #1} (19#2) #3}
 
\def\np#1#2#3{           {\it Nucl. Phys. }{\bf #1} (19#2) #3}
\def\pl#1#2#3{           {\it Phys. Lett. }{\bf #1} (19#2) #3}
\def\pr#1#2#3{           {\it Phys. Rev. }{\bf #1} (19#2) #3}

\def\prl#1#2#3{          {\it Phys. Rev. Lett. }{\bf #1} (19#2) #3}

\def\n.c.#1#2#3{         {\it Nuovo Cim. }{\bf #1} (19#2) #3}
\def\r.n.c.#1#2#3{       {\it Riv. del Nuovo Cim. }{\bf #1} (19#2) #3}

\def\ppnp#1#2#3{           {\it Prog. Part. Nucl. Phys. }{\bf #1} (19#2) #3}

\newcommand{\bi}{\bibitem}
\begin{document}
\thispagestyle{empty}
\begin{titlepage}
\today
\begin{center}
\hfill FTUV/96-07\\
\hfill IFIC/96-08\\
\vskip 0.3cm
\LARGE
{\bf An improved cosmological bound on the tau-neutrino mass}
\end{center}
\normalsize
\vskip 1cm
\begin{center}
{\bf A.D.~Dolgov 
\footnote{Permanent address: ITEP, 113259, Moscow, Russia.}, 
S.~Pastor, and J. W. F. Valle
\footnote{E-mail valle@flamenco.ific.uv.es}\\
}
{\it{Instituto de F\'{\i}sica Corpuscular - C.S.I.C. \\
Departament de F\'{\i}sica Te\`orica, Universitat de Val\`encia \\
46100 Burjassot, Val\`encia, SPAIN}}\\[.15in]
\end{center}

\vskip 1cm
 
\begin{abstract}

We consider the influence of non-equilibrium electronic neutrinos 
(and anti-neutrinos) on the neutron-to-proton ratio. These neutrinos
would come from massive $\nu_\tau$ annihilations
$\bar \nu_\tau \nu_\tau \rightarrow \bar \nu_e \nu_e$. 
For sufficiently large $\nu_\tau$ masses this new effect 
would strongly enhance the (n/p)-ratio, leading to a very
stringent bound on the $\nu_\tau$ mass, even adopting a 
rather weak upper bound on the effective number on 
neutrino species during nucleosynthesis.

\end{abstract}
\vfill
\end{titlepage}
\newpage


The tau-neutrino remains the only one which still can have mass 
in the MeV range. This possibility is theoretically viable, since
it can be unstable on cosmological time scales, thus avoiding the
limits set by the relic density \cite{fae}. It is also quite
interesting experimentally, due to the good prospects for improving 
the \nt mass limits at a B meson or tau-charm factory \cite{jj}. 
In addition, an MeV tau neutrino can be quite interesting 
cosmologically \cite{ma1}. 

The present-day experimental limit on its mass is \cite{eps95}: 
\beq{
m_{\nu_\tau} <  23 \, MeV
\label{mlab}
}\eeq
Primordial nucleosynthesis considerations \cite{sarkar} rule out 
$\nu_\tau$ masses in the range \cite{ckst,dr}:
\beq{
0.5\, MeV <  m_{\nu_\tau} <  35\, MeV
\label{mns}
}\eeq
if $\nu_\tau$ is stable on the nucleosynthesis time scale 
and does not possess new interactions capable of depleting
its density in the cosmic plasma
\footnote{Of course such interactions are needed, at some
level, in order to comply with the limit on the relic neutrino 
density, but the time scale involved is much larger.}.
Calculations of spin-flip processes for the case of a Dirac-type 
$\nu_\tau$ shift the lower bound in this interval down 
to 0.2 MeV \cite{dkr}. Altogether these bounds can be summarised as 
$m_{\nu_\tau}^D < 0.2$ MeV and $m_{\nu_\tau}^M < 0.5 $ MeV, where 
``M'' and ``D'' stands for the Dirac and Majorana mass respectively.

All these bounds are obtained under assumption that the effective 
number of extra neutrino species $\Delta k_\nu$ during nucleosynthesis
cannot be larger than 0.4 or 0.6. Recent contradictory data on the
primordial deuterium abundance \cite{dhigh,dlow} may cast some 
doubts on the validity of this limit (for recent analysis see 
refs.\cite{cst}). In particular, if $\Delta k_\nu = 1$ is allowed, 
there may be an open window for neutrino mass somewhere near 20 MeV. 

In this letter we will show however that this is not the case, 
at least for a Dirac-type neutrino, and that high $\nu_\tau$ mass
values are excluded even when we adopt a much weaker upper bound 
on the extra number of neutrino species. Our result follows from the
consideration of non-equilibrium $\nu_e$'s and $\bar \nu_e$'s
originating from the annihilation of heavy $\nu_\tau$'s. The
effect of non-equilibrium electronic neutrinos on the kinetics 
of (n-p)-reactions was considered in refs.\cite{sato}, for
the case where the $\nu_e$'s and $\bar \nu_e$'s arise from
heavy particle decays during nucleosynthesis. Non-equilibrium
$\nu_e$'s and $\bar\nu_e$'s produced by $e^+e^-$-annihilation 
in the standard model also change the (n/p)-ratio but the effect 
is small \cite{df}. In contrast, the effect we consider here
is much more significant. For example, for the case of a very 
heavy $m \sim 20\,MeV$ $\nu_\tau$ we show that it is equivalent 
to 0.8 extra massless neutrinos species for the case of a Dirac 
$\nu_\tau$, and to 0.1 extra neutrinos for the Majorana case. 
The importance of this effect was mentioned in ref.\cite{dr} 
but the calculations were not performed there because with the 
bound $\Delta k_\nu < 0.6$ the complete range of high $\nu_\tau$ 
mass values was already excluded even without this effect.

One can understand the essential features of the relevant phenomena 
in the following way. At some stage $\nu_\tau$'s, being massive, went 
out of thermal equilibrium. Their number density tends to a ``frozen'' 
value, but their residual annihilation produced energetic non-equilibrium 
$\nu_e$'s and $\bar\nu_e$'s. If this take place sufficiently late, 
roughly speaking, when $T \lsim 2\, MeV$, the produced non-equilibrium
$\nu_e$ and $\bar\nu_e$ are not thermalized. The resulting distortion
of the $\nu_e$ spectrum would shift the frozen (n/p)-ratio and,
correspondingly, the primordial $^4 He$ abundance. The effect has
different signs for high and low energy $\nu_e$'s. If there is an 
excess of $\nu_e$'s (or $\bar\nu_e$'s) at the high energy tail of 
the spectrum, the (n/p)-ratio increases. Indeed, this ratio is 
determined by the competition between the universe expansion 
rate and the rates of the reactions:
\beq{
 p + e ^- \leftrightarrow n + \nu_e
\label{pe}
}\eeq
\beq{
 n + e ^+ \leftrightarrow p + \bar\nu_e
\label{ne}
}\eeq
An excess of energetic $\nu_e$'s and $\bar\nu_e$'s destroys
neutrons in the first reaction and produces them in the second one, 
with the same rate. However, since the number density of protons at 
the moment of freezing of these reactions is about 6 times larger 
than that of neutrons, the second process is more efficient and 
we get more neutrons than in the standard case. 
On the other hand if there is an excess of neutrinos at low energy, 
its effect on the process (\ref{ne}) is small because of the threshold 
effect. However, it enhances the destruction of neutrons in the reaction 
(\ref{pe}) and, correspondingly, the (n/p)-ratio goes down. There is also 
another effect related to the increased overall number density of 
electronic neutrinos coming from the 
$\bar \nu_\tau \nu_\tau \rightarrow \bar \nu_e \nu_e$ annihilation.
This effect decreases the (n/p)-freezing temperature as well as
the (n/p)-ratio independently of the energies of the electronic neutrinos. 
However, as we will see in what follows, the spectrum distortion at high
energies is more important and the net result of all these effects is to
enhance the (n/p)-ratio leading to more stringent \nt mass limits.

Though we obtained our results numerically, an accurate analytical
treatment is possible for large values of the $\nu_\tau$ mass
$m_{\nu\tau} > 10\,MeV$. The kinetic equation which governs the 
spectrum of electronic neutrinos has the form:
\begin{eqnarray}
{\partial f_{e1} \over\partial t} - 
H \omega_1 {\partial f_{e1} \over \partial \omega_1 } = 
{ 1\over2\omega_1 } \int {d^3 p_1 \over (2\pi)^3 2E_1} 
{d^3 p_2 \over (2\pi)^3 2E_2}{d^3 k_2 \over (2\pi)^3 2\omega_2}  
\nonumber \\
(2\pi)^4 \delta^4( p_1+p_2-k_1-k_2 )  
\sum \mid A^2 \mid
( f_{\tau 1} f_{\tau 2} - f_{e 1} f_{e 2}  )
\label{dtf}
\end{eqnarray}
where $f_{\tau}$ and $f_e$ are the $\nu_\tau$ and $\nu_e$ spectral 
functions (or occupation numbers), respectively, and $H$ is the 
Hubble parameter. If the expansion is dominated by photons, $e^-$,
$e^+$, and two massless neutrino species, the effective number of 
degrees of freedom is $g_* = 9$ so that $H \approx 5T^2 /m_{Pl}$.
As a first approximation we take into account only the reaction
$\nu_{\tau 1} + \bar\nu_{\tau 2} \leftrightarrow 
\nu_{e 1}+\bar\nu_{e 2 }$ in the collision integral.
Here $A$ is the amplitude of this reaction and the summation 
is made over the spins of all particles except for $\nu_{e 1}$. 
For the case of a Dirac $\nu_\tau$ one has
\beq{
\sum \mid A^2_D \mid = 32 G^2 _F (p_1 k_1)^2 ,
\label{a2d}
}\eeq
while for the Majorana case:
\beq{
\sum \mid A^2_M \mid = 16 G^2 _F [(p_1 k_1)^2 + (p_2 k_1)^2 
-m^2 (k_1 k_2)]
\label{a2m}
}\eeq
where the identical particle factor $1/2!$ has already been 
included in (\ref{a2m}).

The other relevant processes temporarily omitted in the right 
hand side (r.h.s.) of eq.(\ref{dtf}), namely elastic scattering of 
$\nu_e$ ($\bar \nu_e$) on themselves and other light fermions 
are essential at higher temperatures. They smooth down spectral
distortions and force the distribution back into equilibrium,
somewhat diminishing the effect. We will take them into account 
in what follows (see eqs.(\ref{dxfd}, \ref{dxfm})).

We will solve equation (\ref{dtf}) (or equivalently eqs.(\ref{dxfd}, 
\ref{dxfm})) perturbatively assuming that $f_e$ has the form 
$f_e = f_e ^{(eq)} + \delta f$, where for the equilibrium 
part we take the Boltzmann expression 
\mbox{$ f_e ^{(eq)}= \exp (-\omega/T)$}. 
Quantum statistics corrections may diminish the 
effect by (5-10)\% \cite{dk}. We neglect terms of the second order in 
$\delta f$. The distortion of the electronic neutrino spectrum
is induced by the deviation of the massive $\nu_\tau$ from thermal
equilibrium which arises when the temperature of the cosmic plasma
drops below the $\nu_\tau$ mass. It is usually assumed that 
kinetic equilibrium is maintained, while an effective chemical
potential of the same magnitude for particles and antiparticles
is generated \cite{adbar}. In other words $f_\tau$ takes the form 
$f_\tau = \exp (\xi(t) - E/T)$. The $\nu_\tau$ number density 
is calculated from the well known equation:
\beq{
\dot n_\tau + 3Hn_\tau = \langle \sigma v \rangle 
(n^{(eq) 2}_\tau - n^2_\tau )
\label{dtntau}
}\eeq
where $\langle \sigma v \rangle$ is the thermally averaged 
cross-section of $\nu_\tau$-annihilation (multiplied by velocity)
and $n_\tau^{(eq)}$ is their equilibrium number density. We 
write $n_\tau$ as usual, in the form $ n_\tau = r n_0$ where 
$n_0 \approx 0.09 g_s T^3$ is the equilibrium number density of
massless neutrinos with the same temperature and $g_s$ is the spin
counting factor. We have calculated $r$ numerically solving 
eq.(\ref{dtntau}) and parametrised the solution in the form 
$r=r_0+r_1m_{\nu_\tau}/T$. The results we obtain agree 
with those of ref. \cite{dr}.

We assume that the temperature drops in the usual way, 
$\dot T = -HT$, possible corrections to this law are not essential
here. In this case the equilibrium part of $f_e$ annihilates the 
r.h.s. of eq.(\ref{dtf}) and it would contain only $\delta f$. 
Introducing new variables $x \equiv m/T$ and $y \equiv \omega/T$ 
we can rewrite equation (\ref{dtf}) in the form:
\beq{
Hx {\partial \delta f_D \over \partial x} = {2G_F^2 m^5 \over 3\pi^3}
\left({n_\tau^2  \over n_\tau^{(eq) 2}} -1 \right)
{ \exp(-y-\alpha) \over x^2 y^{1/2} } \int^\infty_0 du e^{-u}
\left( {3\over 4} + {u\over \alpha}\right) 
\sqrt{u \left( 1+ {u \over \alpha} \right) } + ...    
\label{dxfd}
}\eeq
for the Dirac case while, for the Majorana case we obtain
\beq{
Hx {\partial \delta f_M \over \partial x} = {2G_F^2 m^5 \over 3\pi^3}
\left( {n_\tau^2 \over n_\tau^{(eq) 2}}-1 \right)
{ \exp(-y-\alpha) \over x^4 y^{-1/2} } \int^\infty_0 du e^{-u}
u^{3/2} \sqrt { 1+ {u \over \alpha} } + ...    
\label{dxfm}
}\eeq
Here $\alpha \equiv x^2/y$ and the dots denote contributions 
from elastic scattering which we have so far omitted.

In the limit of high $\nu_\tau$ mass, i.e. for $x \gg 1$ and 
$\alpha \gg 1$, the integrals can be performed approximately 
so that the r.h.s. of these equations can be written as:
\beq{
{\partial \delta f \over \partial x} \approx 0.96 \times 10^{-2} m^3 
{r^2 - r^{(eq) 2} \over x^4 y^{1/2} } 
\exp \left[-{( x-y )^2 \over y}\right] F_D(x,y)+...
\label{dxfd1}
}\eeq
(Dirac case),
\beq{
{\partial \delta f \over \partial x} \approx 1.92 \times 10^{-2} m^3 
{r^2 - r^{(eq) 2} \over x^6 y^{-1/2} } 
\exp \left[ -{( x-y )^2 \over y}\right] F_M(x,y)+...
\label{dxfm1}
}\eeq
(Majorana case).

Here and in what follows $m$ is measured in MeV,
$F_D=(1+11y/4x^2+65y^2/32x^4)/(1+15/4x+165/32x^2)$,
and $F_M=(1+5y/4x^2-35y^2/32x^4)/(1+15/4x+165/32x^2)$. The numerators
come from the expansion of the integrals in the r.h.s. of equations
(\ref{dxfd}, \ref{dxfm}) and the denominators come from the 
non-relativistic expansion of the number density.

It is straightforward to take into account the other processes where
$\nu_e$ and $\bar \nu_e$ participate, denoted by ``...'' in the above
equations. They are either elastic scattering off light leptons, 
$ \nu_e + l \leftrightarrow \nu_e + l$, or 
$ \nu_e + \bar \nu_e \leftrightarrow l + \bar l$ annihilations.
These processes tend to restore equilibrium, thus diminishing the 
distortion $\delta f$. The collision integral in the r.h.s. of eq.
(\ref{dtf}) contains $\delta f$ in terms of two types: first, 
when $\delta f = \delta f (k_1)$ refers to the particle under 
consideration and, second, when $\delta f = \delta f (k_2)$ 
refers to one over whose momentum we integrate in $d^3 k_2$. 
The first ones come with a negative sign and force the system 
into equilibrium, while the terms of the second type typically 
appear with positive sign and induce a deviation from equilibrium
\footnote{The terms of the second kind may also appear with a negative 
sign, like e.g. in the process $\nu_e + \nu_e \leftrightarrow \nu_e +\nu_e$
from neutrinos in the initial state, but the contribution from similar terms 
coming from neutrinos in the final state over-weights the negative one.}.
In order to account for the terms of the second kind one has to solve 
a complicated integro-differential kinetic equation. In contrast 
it is very easy to treat the terms of the first kind. They result 
in the addition of an extra term of the form $(-\delta f) F(x,y)$,
with a known function $F(x,y)$, to the r.h.s. of eqs.(\ref{dxfd},\ref{dxfm}). 
In what follows we take into account only the contribution of the 
terms of the first kind, thus somewhat overestimating the suppression 
factor. The real effect should be somewhat larger. However, even in this 
approximation our effect results in a considerable improvement
of the strength of the bound on $m_{\nu_\tau}$. 

Taking into account all possible reactions of $\nu_e$ with light
leptons we get $F\approx m^3y/3x^4$. The result is valid for arbitrary 
$x$ and $y$ values (the restriction $x\gg 1$ is not assumed here).
After a simple algebra we get the following equations for the 
spectrum distortion:    
\beq{
{\partial \delta f_D \over \partial x} = 0.0096 m^3 
{ r^2 - r^{(eq) 2} \over x^4 y^{1/2} } 
\exp \left(-{(x-y)^2 \over y} \right) F_D(x,y)
- {m^3 y \over 3 x^4} \delta f_D 
\label{dxfd2}
}\eeq
\beq{
{\partial \delta f_M \over \partial x} = 0.0192 m^3 
{ r^2 - r^{(eq) 2} \over x^6 y^{-1/2} } 
\exp \left(-{(x-y)^2 \over y} \right) F_M(x,y)
- {m^3 y \over 3 x^4} \delta f_M 
\label{dxfm2}
}\eeq
where ``M'' and ``D'' stand for the Dirac and Majorana cases, respectively.

The solutions of these equations can be easily written. These 
expressions should be substituted into the equation which governs 
the neutron-to-proton ratio. Since the spectrum is distorted at high 
neutrino energies we may neglect the electron mass, leading to 
the approximate form (see e.g. ref.\cite{df}): 
\begin{eqnarray}
{dr_n \over dT} = -0.05 T^2 \Bigl [ (24+12\beta + 2\beta^2) 
(e^{-\beta}-r_n (1+ e^{-\beta})) + \nonumber \\
(1/2)\int_0^\infty dy y^2 (y+\beta)^2 (\delta f (y+\beta) -
r_n ( \delta f (y) + \delta f (y+\beta) )) \Bigr ]
\label{drndt}
\end{eqnarray}
Here $r_n$ is the ratio of the neutron number density to the
total baryonic number density. The (n/p)-ratio is expressed 
through $(n/p)=r_n/(1-r_n)$, and $\beta \equiv \Delta m /T$
where $\Delta m = 1.3$ MeV is the neutron-proton mass difference.
Of course, in our numerical integration we have used the exact 
equation for the (n/p)-ratio as given e.g. in ref.\cite{df}.

We can now roughly estimate the variation of (n/p)-ratio using 
eq.(\ref{drndt}) in the following way. First, we integrate
eqs. (\ref{dxfd2},\ref{dxfm2}) neglecting the last term in 
the r.h.s. In this approximation the $\nu_e$'s and 
$\bar \nu_e$'s are in thermal equilibrium above $T \sim 2$ MeV,
while below this temperature the elastic scattering processes are not 
important. We estimate the relevant integral over $x$ in the
Gaussian approximation near $x=y$ and integrate the obtained 
result over $y$ as written in eq.(\ref{drndt}). So this equation takes
the form:
\begin{eqnarray}
{dr_n \over dT} = -0.05T^2 \Bigl [(24+12\beta+2\beta^2)
e^{-\beta}+\Delta^{-}/2 - \nonumber \\
r_n((24+12\beta+2\beta^2)(1+e^{-\beta}) + 
\Delta^{+}/2 + \Delta^{-}/2)\Bigr ]
\label{drdel}
\end{eqnarray}
where
\beq{
\Delta^{\pm}_D \approx 0.017 r_D^2 m^3 x \left( 1-{x_i\over x} \pm
{2\Delta m \over m}\ln {x \over x_i} \right )
\label{deltad}
}\eeq

\beq{
\Delta^{\pm}_M \approx 0.034 r_M^2 m^3  \left( 
\ln {x \over x_i} \pm 2\beta \left({1\over x_i} - 
{1\over x}\right ) \right )
\label{deltam}
}\eeq
with $x_i=m/T_i=m/2$. The signs ``+'' and ``$-$'' 
in the above expressions correspond respectively to the integrals 
$\int dy y^2 (y+\beta)^2 \delta f(y)$ and to
$\int dy y^2 (y+\beta)^2 \delta f(y+\beta)$.
 
As we have already noted, the modification of the neutrino
spectrum results in two opposite effects. First, the 
``equilibrium'' value of $r_n$ (obtained by equating 
the r.h.s. to zero) increases. Second, the (n-p)-reactions get
frozen later, as the excess of energetic $\nu_e$'s and 
$\bar \nu_e$'s increases their rates. Taking into account 
both effects we can write the correction to $r_n$ as: 
\beq{
{\delta r_n \over r_n} = 
\Bigl [1+\frac{\Delta^{-}e^{\beta_1}-
(\Delta^{+}+\Delta^{-})/(1+ e^{-\beta_1})}
{48+24\beta_1+4\beta_1^2} \Bigr ]
e^{\beta_0-\beta_1}\frac{1+ e^{-\beta_0}}{1+ e^{-\beta_1}}
-1
\label{deltar}
}\eeq
where $\beta_0 \equiv \Delta m /T_{f0}$, $T_{f0}=0.65$ MeV
is the standard value of (n-p)-freezing temperature
and $\Delta^{\pm}$ are taken at $x=m/T_{f0}$. The new
freezing temperature $T_{f1}$ can be found from
\beq{
\beta_1 = \beta_0 + \frac{\beta_0^4}
{(48+24\beta_0+4\beta_0^4)(1+ e^{-\beta_0})}
\int^\infty_{\beta_0} dw~\frac{\Delta^{+}+\Delta^{-}}{w^4}
}\eeq
Using these expressions 
we obtain $\delta r_n /r_n = (5-10)\%$ for a Dirac $\nu_\tau$ with 
mass in the range $m=10-20$ MeV, while for the Majorana case
we find $\delta r_n /r_n = (1-2)\%$. These results are in 
reasonable agreement with our numerical calculations
described below.

Now let us briefly describe the calculations. The solution of
eqs.(\ref{dxfd2},\ref{dxfm2}) was taken analytically in terms of 
one-dimensional integral $\int^x_0 dx'$. Then we substituted this
expression into the exact equation governing the (n/p)-ratio. Integration
over $y$ in this equation can be approximately done analytically with
a very good precision. The last integral over $x'$ was performed numerically
for different values of $x$ and accurate interpolating functions were
substituted into the equation for the (n/p)-ratio. This first order 
differential equation was solved numerically.

In the case of Dirac $\nu_\tau$ we obtained $\delta r_n/r_n =4.2\%$
for $m=10$ MeV and 3.6\% for $m=20$ MeV. In the Majorana case we got
1.2\% for 10 MeV and 0.5\% for 20 MeV. These results can be rewritten
as corrections to the effective neutrino number $N_{eq}$ as a function 
of the \nt mass as shown in the figure. The solid curves denote the
results we obtain including the effect of non-equilibrium electronic 
neutrinos and antineutrinos from massive $\nu_\tau$ annihilations
on the neutron-to-proton ratio. The dashed curves represent the
corresponding results neglecting these corrections \cite{ckst,dr}.
As can be seen from the figure the increase of $N_{eq}$ in the 
mass range of interest is larger for the case of Dirac than it is 
for the Majorana neutrino case. From our results one concludes
that relatively long-lived massive tau neutrinos above the few 
MeV range are ruled out by the requirement $N_{eq} \leq 4$.
Thus it would seem from our results that the only way to 
accommodate a massive tau neutrino above the few MeV range 
is if there are new interactions beyond those of the standard model
that can make it decay and/or annihilate efficiently on the 
nucleosynthesis time scale. The case of unstable $\nu_\tau$'s 
has been considered in refs.\cite{unstable}. The effect we have 
discussed would also improve the bounds obtained in these papers.
The alternative case of neutrinos with large majoron
annihilation cross sections is under consideration \cite{dprv}.

As this work was completed we became aware of the paper \cite{fko}
where the non-equilibrium heating of electronic neutrinos by 
$\nu_\tau \bar \nu_\tau$-annihilation was considered. The conclusion of
this paper about the influence of non-equilibrium $\nu_e$'s and 
$\bar \nu_e$'s on the (n/p)-ratio is opposite to ours. We think 
that the results differ because these authors took into account 
only a decrease of the (n-p)-freezing temperature 
due to the excess of electronic neutrinos, but neglected the 
spectrum distortion which we have discussed here and which
turns out to have a stronger impact on the (n/p)-ratio. 


This work was supported by DGICYT under grants
PB92-0084 and SAB94-0089 (A. D.). S.P. was supported by
Conselleria d'Educaci\'o i Ci\`encia of Generalitat Valenciana.

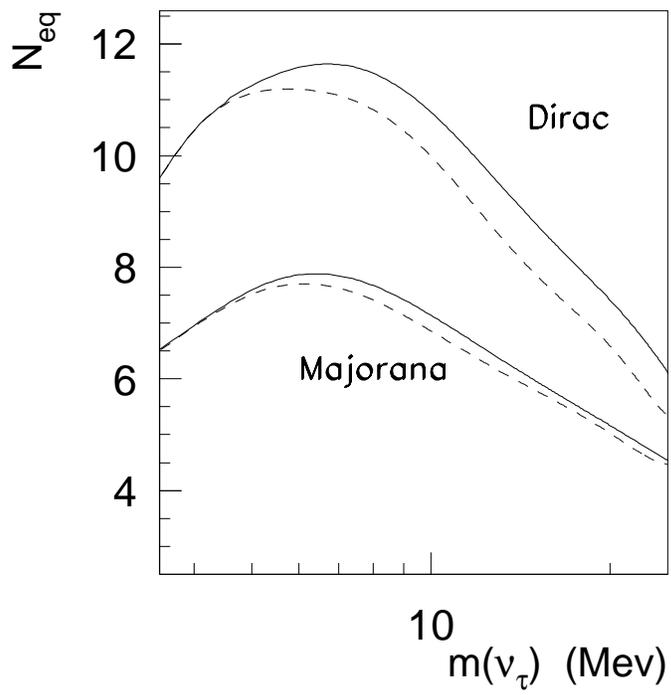
\end{document}